# Energy Optimization in Massive MIMO UAV-Aided MEC-Enabled Vehicular Networks

Emmanouel T. Michailidis, Nikolaos I. Miridakis, *Senior Member, IEEE*, Angelos Michalas, Emmanouil Skondras, Dimitrios J. Vergados, and Dimitrios D. Vergados

*Abstract*—This paper presents a novel unmanned aerial vehicle (UAV)-aided mobile edge computing (MEC) architecture for vehicular networks. It is considered that the vehicles should complete latency-critical computation-intensive tasks either locally with on-board computation units or by offloading part of their tasks to road side units (RSUs) with collocated MEC servers. In this direction, a hovering UAV can serve as an aerial RSU (ARSU) for task processing or act as an aerial relay and further offload the computation tasks to a ground RSU (GRSU). In order to significantly reduce the delay during data offloading and downloading, this architecture relies on the benefits of massive multiple-input–multiple-output (MIMO). Therefore, it is considered that the vehicles, the ARSU, and the GRSU employ large-scale antennas. A three-dimensional (3-D) geometrical representation of the MEC-enabled network is introduced and an optimization method is proposed that minimizes the weighted total energy consumption (WTEC) of the vehicles and ARSU subject to transmit power allocation, task allocation, and timeslot scheduling. The numerical results verify the theoretical derivations, emphasize on the effectiveness of the massive MIMO transmission, and provide useful engineering insights.

## I. Introduction

WITH the emergence of the big data era at vehicular networks, the Internet of Vehicles (IoV) paradigm, and the vehicular-to-everything (V2X) information interaction, a vast number of connected automobile terminals equipped with computation and multi-communication units will pave the path for novel services [1]. The next wave of applications, including augmented reality (AR), ultra-high-quality video streaming, and autonomous driving, is expected to reach the limits of current technologies and pose strict requirements in terms of computation, latency, and throughput. For locally intra-vehicle processed applications, a large amount of energy is consumed, which in turn reduces the driving range of energy-limited electric vehicles [2]. On the other hand, it is often infeasible for resource-constrained vehicles to timely handle computation-intensive tasks. To maintain the energy consumption at a low level, while meeting critical latency demands, partly or fully task offloading to mobile edge computing (MEC) servers has been suggested [3], [4]. In this direction, road side units (RSUs) along roads and in proximity to the vehicles can facilitate the provision of MEC services.

### A. Background

As the highly dynamic and random nature of MEC-enabled vehicular networks drastically affects the performance of data offloading, the computation overhead was minimized in [5], whereas a reliability-oriented stochastic optimization model was presented in [6] to maximize the lower bound of the expected reliability during offloading. The effective collaboration of cloud computing and MEC in vehicular networks was addressed in [7] and the computation offloading decisions were optimized through a game-theoretic approach. Also, a vehicle edge computing (VEC) network architecture with the vehicles acting as MEC servers was proposed in [8] and the joint optimization of computing offloading and resource allocation was handled via deep reinforcement learning (DRL). A software-defined networking (SDN)-based and fiber-wireless (FiWi)-enhanced load-balancing task offloading policy was introduced in [9] to minimize the processing delay. Moreover, a computation offloading protocol that exploits geo-location information was developed in [10] enabling efficient and reliable data retrieval in VEC environments with hybrid vehicle-to-vehicle (V2V) and vehicle-to-infrastructure (V2I) links. In order to minimize the total network delay, an edge intelligence empowered IoV framework was constructed in [11] and an online algorithm relying on Lyapunov optimization was designed to manage computation offloading and content caching.

Despite such promising computing capabilities, attaining ubiquitous connectivity and sufficient radio coverage between vehicles and MEC servers is challenging, since ground RSUs (GRSUs) often struggle in areas with obstacles and highly mobile and disperse nodes. In this regard, hovering aerial RSUs (ARSUs) based on unmanned aerial vehicles (UAVs)

E. T. Michailidis is with the Department of Electrical and Electronics Engineering, University of West Attica, Egaleo, 12241, Greece (e-mail: emichail@uniwa.gr).
N. I. Miridakis is with the Department of Informatics and Computer Engineering, University of West Attica, Egaleo, 12243, Greece (e-mail: nikozm@uniwa.gr).
A. Michalas is with the Department of Electrical and Computer Engineering, University of Western Macedonia, Kozani, 50131, Greece (e-mail: amichalas@uowm.gr).
E. Skondras is with the Department of Informatics, University of Piraeus, Piraeus, 18534, Greece (e-mail: skondras@unipi.gr).
D. J. Vergados is with the Department of Informatics, University of Western Macedonia, Kastoria, 52100, Greece (e-mail: dvergados@uowm.gr).
D. D. Vergados is with the Department of Informatics, University of Piraeus, Piraeus, 18534, Greece (e-mail: vergados@unipi.gr).

can fly over connected vehicles and effectively mitigate shadowing and blockage effects thus maintaining line-of-sight (LoS) propagation [12]-[14]. Most of current work on UAV-aided MEC-enabled networks has focused on energy-aware solutions both from ground users (GUs) and UAV perspective. In [15], a UAV was deployed to assist an access point (AP) to provide MEC services to GUs and an algorithm that minimizes the energy consumption was proposed. By adopting similar setups, the maximum delay [16], sum power [17], task completion time [18], average latency [19], and computation efficiency [20] were also optimized. A non-orthogonal multiple access (NOMA) scheme was studied in [21], whereas an edge-cloud system supporting virtualized network functions (VNFs) was proposed in [22]. Beyond the deterministic binary and partial task offloading, the concept of stochastic offloading was studied in [23]. In [24], the resource allocation and UAV's trajectory were optimized for a social IoV (SIoV) scenario, whereas an SDN-enabled computation offloading optimization framework for vehicular networks was proposed in [25] to minimize the execution time of the computation tasks of vehicles, under energy and quality of service (QoS) constraints. The advantage of employing UAVs as MEC servers in Cyber-Physical Systems (CPSs) was outlined in [26] and the three-dimensional (3-D) UAV's trajectory was sub-optimally optimized to extend the UAV's endurance. In [27], the use of UAV-mounted edge nodes in Long Range Wide Area Networks (LoRaWANs) for disaster management was investigated. Wireless power transfer (WPT) was also introduced to prolong network's operation time. In this respect, optimization problems were formulated to maximize the sum completed task-input bits [28] and the UAV's required energy [29]. An Internet of Things (IoT) scenario was examined in [30], while a multi-UAV-based MEC system was proposed in [31].

On another front, massive multiple-input–multiple-output (MIMO) technology has recently received unprecedented attention as a key enabler for increased spectral and energy efficiency, drastically reduced round-trip latency, and support of highly-intensive computation tasks for a large number of connected users [32]. Previous work has investigated the adoption of massive MIMO in vehicular (e.g. [33]) and UAV-based (e.g. [34], [35]) scenarios, without focusing on MEC applications. To the best of the authors' knowledge, the area of massive MIMO UAV-based MEC networks is unexplored and only single-antenna solutions have been previously studied. However, these solutions cannot properly capture the massive MIMO channel characteristics. In the context of terrestrial MIMO MEC networks, a multi-antenna NOMA architecture that enables multi-user computation offloading over the same time/frequency resources was proposed in [36]. Also, the optimization of energy consumption and maximum delay, under perfect and imperfect channel state information (CSI) estimation, was studied for MIMO [37] and massive MIMO [38] systems. Moreover, single-cell [39] and multi-cell [40] MEC networks that enable the simultaneous offloading of multiple APs were previously presented. The benefits of combining massive MIMO and millimeter wave (mmWave) frequencies in wireless local area networks (WLANs) with MEC were underlined in [41], whereas a cell-free system consisting of multiple single/multi-antenna APs with MEC servers and a central cloud server was described in [42]. Notwithstanding, these works are improper for UAV-based MEC networks, since the UAVs fly in a 3-D space and above rooftops leading to peculiar link geometry and especial mobility characteristics. To reconcile these challenging issues, newer network architectures are indispensable.

*B. Contribution*

Motivated by the aforementioned observations, we investigate a massive MIMO UAV-aided MEC-enabled vehicular network. The major contributions of this paper are summarized as follows:

- **A novel dual-MEC network architecture** is proposed, where a UAV operates as an ARSU equipped with a MEC server and also as an intermediate decode-and-forward (DF) aerial relay between vehicles and an GRSU. This architecture trades on the massive MIMO transmission and the efficient use of all computing resources. Moreover, a MEC computation offloading and downloading protocol is presented with distinct operation phases. Based on this protocol, partial offloading is applied to obtain a trade-off between energy consumption and delay.

- In order to unlock the full potential of massive MIMO, we propose **the concept of triple-sided massive MIMO**, as an extension of the single-sided [34], [43] and double-sided massive MIMO [35]. Therefore, it is considered that the vehicles, the ARSU, and the GRSU employ two-dimensional (2-D) uniform rectangular planar arrays (URPAs) with a large number of antenna elements. Although there exist some practical barriers (e.g., power consumption and complexity) towards the implementation of large-scale antennas, we believe that these issues will be handed in the near future and practical transceivers will be developed [44].

- **Realistic 3-D placement and mobility modeling** of the vehicles, the ARSU, the GRSU, and the URPAs is proposed. Also, position, distance, and velocity vectors are used to accommodate the geometrical representation of the proposed network architecture and construct the massive MIMO channel matrices.

- **A multi-variable optimization problem is formulated** that intends to minimize the weighted total energy consumption (WTEC) from both the vehicles and ARSU perspective and prolong their lifetime as major network segments. The Lagrange dual method is leveraged to derive closed-form solutions for the transmit power allocation, time slot scheduling, and computation bits allocation. A subgradient-based algorithm is also constructed to expedite the optimization process. The results depict the total computation-based and communication-based delay (TCCD) and WTEC, point out the advantages of massive MIMO transmission, and validate the effectiveness of the optimization procedure.

## C. Structure

The remainder of this paper is organized as follows. Section II introduces the system model and outlines the geometrical, mobility, and channel characteristics. Section III presents the computation offloading and downloading model. Section IV formulates the optimization problem and derives its solution, whereas Section V provides numerical results. Finally, conclusions and future directions are given in Section VI.

## II. SYSTEM MODEL

Consider a triple-sided massive MIMO MEC-enabled vehicular network that facilitates the computing offloading of $K$ vehicles moving along a unidirectional road segment. Each vehicle has a latency-sensitive and bit-wise-independent computation task that can be executed partly *locally* with the on-board computing processor and partly *remotely* by computation offloading to network MEC servers. In this direction, a fixed GRSU with sufficiently powerful computation capacity and grid power supply is situated along the road. It is assumed that the $k$-th vehicle cannot directly communicate with the GRSU owing to signal blockage or severe shadowing. Thus, an ARSU is employed to enable vehicle-to-ARSU (V2U) networking facilitating the MEC services. Contrary to GRSU, the ARSU has certain computing and energy limitations that depend on its type, weight, and battery size. To preserve its energy resources, the ARSU can determine the portion of tasks that can locally process and then act as an aerial relay forwarding the remaining part of the vehicles' offloaded tasks to GRSU. By employing a sufficiently large data buffer, the ARSU can separately store the offloaded data and the computation results.

### A. Geometrical Characteristics and Mobility Model

Fig. 1 illustrates the 3-D geometrical characteristics of the proposed dual-MEC network architecture. To aid our analysis, the subscripts $k$, $U$, and $R$, where $k \in \{1,2,...K\}$, are associated with the $k$-th vehicle, ARSU, and GRSU, respectively. The ($x$, $y$, $z$) axes designate the global coordinate system (GCS), which controls the position of each network segment, with the projection $\tilde{O}_U$ of ARSU's array center $O_U$ positioned at the origin ($x = 0$, $y = 0$, $z = 0$), where $\tilde{O}$ denotes the projection of a point $O$ onto the $xy$ plane. Since the height of $k$-th vehicle (GRSU) is relatively low compared to that of ARSU, $O_k$ ($O_R$) is almost identical to $\tilde{O}_k$ ($\tilde{O}_R$). It is considered that the $k$-th vehicle, ARSU, and GRSU employ uniform rectangular planar arrays (URPAs). The URPAs are defined by the local coordinate system (LCS), the origins of which are at URPAs' centers. More specifically, the $k$-th vehicle is equipped with an URPA with $L_k = L_{kx} \times L_{ky}$ antennas spanning $L_{kx}$ rows along the $x$-axis and $L_{ky}$ columns along the $y$-axis of the LCS with equal inter-element spacing $\delta_k$, where $L_k \gg 1$. The position of URPAs in the GCS is specified by the transformation from GCS to LCS, whereas the angles $\beta_x \in [-\pi/2, \pi/2]$ (slant angle), $\beta_y \in [-\pi/2, \pi/2]$ (downtilt angle), and $\beta_z \in [0, 2\pi]$ (bearing angle) designate a 3-D counterclockwise rotation of LCS with reference to GCS and can also describe ARSU's roll, pitch, and yaw, respectively [43]. Then, a sequence of rotations assigns the URPAs' orientation. For the position vector of antenna element $A_{mm'}^k$ with $m = 1,2,...,L_{kx}$, $m' = 1,2,...,L_{ky}$, we obtain $\mathbf{A}_{mm'}^k = \mathbf{R}\mathbf{A}_{mm'}^{\prime k}$, where

$$\mathbf{R} = \mathbf{R}_X(\beta_x)\mathbf{R}_Y(\beta_y)\mathbf{R}_Z(\beta_z) = \begin{pmatrix} 1 & 0 & 0 \\ 0 & \cos\beta_x & -\sin\beta_x \\ 0 & \sin\beta_x & \cos\beta_x \end{pmatrix}$$

$$\times \begin{pmatrix} \cos\beta_y & 0 & \sin\beta_y \\ 0 & 1 & 0 \\ -\sin\beta_y & 0 & \cos\beta_y \end{pmatrix} \begin{pmatrix} \cos\beta_z & -\sin\beta_z & 0 \\ \sin\beta_z & \cos\beta_z & 0 \\ 0 & 0 & 1 \end{pmatrix}, \quad (1)$$

$\mathbf{A}_{mm'}^{\prime k} = \left[ x_{mm'}^{\prime k}, y_{mm'}^{\prime k}, 0 \right]^T$ is the antenna vector in the LCS,

$$x_{mm'}^{\prime k} = \begin{cases} -(L_{kx} - 2m + 1)\delta_k / 2, & m < (L_{kx} + 1)/2 \\ (L_{kx} - 2m + 1)\delta_k / 2, & m \geq (L_{kx} + 1)/2 \end{cases}, \quad (2)$$

$$y_{mm'}^{\prime k} = \begin{cases} -(L_{ky} - 2m' + 1)\delta_k / 2, & m' < (L_{ky} + 1)/2 \\ (L_{ky} - 2m' + 1)\delta_k / 2, & m' \geq (L_{ky} + 1)/2 \end{cases}, \quad (3)$$

and $(\cdot)^T$ denotes the transpose operation. By replacing the indices, the position vectors $\mathbf{A}_{pp'}^U$ and $\mathbf{A}_{qq'}^R$ of antenna elements $A_{pp'}^A$ and $A_{qq'}^R$, respectively, are similarly defined.

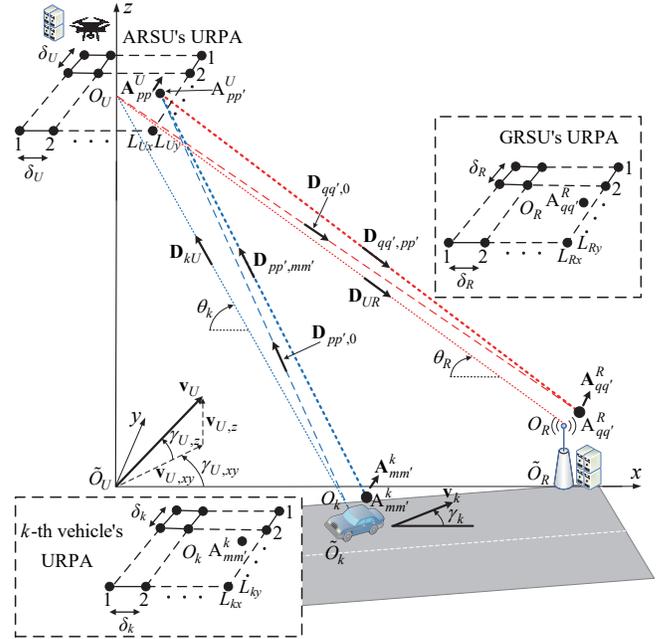

Fig. 1. The geometrical characteristics of the proposed triple-sided massive MIMO UAV-aided MEC-enabled vehicular network, where an ARSU assists the $k$-th vehicle execute its offloaded computing task and also acts as an aerial relay to further transmit part of this task to an GRSU for computing.

Based on Fig. 1, $\mathbf{D}_{pp',mm'} = \mathbf{D}_{pp',0} - \mathbf{A}_{mm'}^k$ is the distance vector between $\mathbf{A}_{mm'}^k$ and $\mathbf{A}_{pp'}^U$, $\mathbf{D}_{pp',0} = \mathbf{D}_{kU} + \mathbf{A}_{pp'}^U$ is the distance vector between the $k$-vehicle array center and $\mathbf{A}_{pp'}^U$, $\mathbf{D}_{kU} = [h_U/\tan\theta_k, 0, h_U]^T$ is the distance vector between the $k$-vehicle and ARSU array centers, $h_U$ is the altitude of the ARSU antenna array, and $\theta_k$ is the elevation angle of ARSU relative to $O_k$. The other distance vectors can be similarly defined. Moreover, $\mathbf{v}_k = v_k[\cos\gamma_k, \sin\gamma_k, 0]^T$ and $\mathbf{v}_U = v_U[\cos\gamma_{U,xy}\cos\gamma_{U,z}, \sin\gamma_{U,xy}\cos\gamma_{U,z}, \sin\gamma_{U,z}]^T$ denote the velocity vectors of the $k$-vehicle and the ARSU, respectively, $v_k(v_U)$ is the instantaneous velocity of the $k$-vehicle (ARSU), $\gamma_k(\gamma_{U,xy})$ is the azimuth angle that controls the moving direction of the $k$-vehicle (ARSU) in the azimuth domain, and $\gamma_{U,z}$ is the elevation angle that characterizes possible rising, diving, and hovering operations of the ARSU.

The total energy of ARSU is limited. Thus, the flying period is restrained by $T_U$. For convenience, adequately small constant $\tau$ is used to divide $T_U$ into $N$ timeslots. In each timeslot $n \in \{1,2,...N\}$, the $k$-th vehicle and the ARSU can be considered to be static, whereas their antenna position vectors are updated, respectively, as $\mathbf{A}_{mm'}^k[n+1] = \mathbf{A}_{mm'}^k[n] + \mathbf{v}_k[n]\tau$ and $\mathbf{A}_{pp'}^U[n+1] = \mathbf{A}_{pp'}^U[n] + \mathbf{v}_U[n]\tau$ with $\mathbf{A}_{mm'}^k(t) = \mathbf{A}_{mm'}^k(\tau n) = \mathbf{A}_{mm'}^k[n]$ and $\mathbf{A}_{pp'}^U(t) = \mathbf{A}_{pp'}^U(\tau n) = \mathbf{A}_{pp'}^U[n]$. Note that the distance vectors are also updated accordingly.

Based on [26], we model the energy consumption during flight in the $n$-th timeslot for a fixed-wing ARSU as

$$E_{fl,fw}[n] = \tau\left(c_1\|\mathbf{v}_{U,xy}[n]\|^3 + \frac{c_2}{\|\mathbf{v}_{U,xy}[n]\|} + c_3\|\mathbf{v}_{U,z}[n]\|\right), \quad (4)$$

where $c_1$ and $c_2$ are constants depending on the ARSU's weight, wing area, and air density, $c_3$ is a constant associated with ARSU's descending/ascending, $\mathbf{v}_{U,xy}[n]$ and $\mathbf{v}_{U,z}[n]$ are the horizontal and vertical ARSU velocity vector, respectively, with $\mathbf{v}_U[n] = \mathbf{v}_{U,xy}[n] + \mathbf{v}_{U,z}[n]$, and $\|\cdot\|$ is the Euclidean norm. For a rotary-wing ARSU, the energy consumption during flight can be modelled as [26]

$$E_{fl,rw}[n] = \tau\left(P_0\left(1 + \frac{3\|\mathbf{v}_{U,xy}[n]\|^2}{v_{tip}^2}\right) + \frac{1}{2}d_r s\rho G\|\mathbf{v}_{U,xy}[n]\|^3 \right.$$
$$\left. + P_1\sqrt{\sqrt{1 + \frac{\|\mathbf{v}_{U,xy}[n]\|^4}{4v_0^2}} - \frac{\|\mathbf{v}_{U,xy}[n]\|^2}{2v_0^2}} + P_2\|\mathbf{v}_{U,z}[n]\|\right), \quad (5)$$

where $P_0$ and $P_1$ describe the blade profile power and induced power, respectively, $P_2$ controls the descending/ascending power, $v_{tip}$ is the tip speed of rotor blade, $v_0$ is the mean rotor induced velocity, $d_r$ is the fuselage drag ratio, $s$ is the rotor solidity, $\rho$ is the air density, and $G$ is the rotor disc area.

### B. Wireless Transmission Model

In each timeslot the $k$-th vehicle and the ARSU move over a small distance. Thus, the channel coefficients are keeping unchanged and can be estimated using uplink and downlink orthogonal pilot sequences (prior to data transmission) at the start of each timeslot [45]. Overall, the channel is described by a series of channel snapshots for different placement of the $k$-th vehicle and the ARSU in each timeslot. It is assumed that the V2U, ARSU to GRSU (U2R), GRSU to ARSU (R2U), and ARSU to $k$-vehicle (U2V) channels are dominated by LoS links within the short frame $T_U$, as also indicated by recent measurements in several propagation environments [46]-[49]. Based on [35], [43], the channel coefficient between antenna elements $\mathbf{A}_{mm'}^k$ and $\mathbf{A}_{pp'}^U$ can be written

$$g_{pp',mm'}[n] = \sqrt{\beta_{kU}[n]} h_{pp',mm'}[n], \quad (6)$$

where

$$\beta_{kU}[n] = \beta_0 \|\mathbf{D}_{kU}[n]\|^{-\alpha}, \quad (7)$$

$$h_{pp',mm'}[n] = \exp\left[j\left(2\pi f_{pp',mm'}[n] + \varphi_{pp',mm'}[n]\right)\right], \quad (8)$$

$$f_{pp',mm'}[n] = \frac{\langle \mathbf{D}_{kU}[n] + \mathbf{A}_{pp'}^U[n] - \mathbf{A}_{mm'}^k[n], \mathbf{v}_k[n] - \mathbf{v}_U[n]\rangle}{\lambda\|\mathbf{D}_{kU}[n] + \mathbf{A}_{pp'}^U[n] - \mathbf{A}_{mm'}^k[n]\|}, \quad (9)$$

$$\varphi_{pp',mm'}[n] = \frac{2\pi\|\mathbf{D}_{kU}[n] + \mathbf{A}_{pp'}^U[n] - \mathbf{A}_{mm'}^k[n]\|}{\lambda}, \quad (10)$$

$\beta_0$ is the channel gain at a reference distance $d_0 = 1$ m, $\alpha$ is the path-loss exponent, $\langle\cdot,\cdot\rangle$ is the inner product operator, and $\lambda$ is the carrier wavelength. Since $\|\mathbf{D}_{kU}[n]\| \gg \{\delta_k, \delta_U\}$, the path-loss parameter $\beta_{kU}[n]$ is invariable over $L_k$ and $L_U$. The massive MIMO channel between the $k$-vehicle and the ARSU in the $n$-th time slot can be described by the matrix $\mathbf{G}_{kU}[n] = [g_{pp',mm'}[n]]_{L_U \times L_k} \in \mathbb{C}^{L_U \times L_k}$, which can be expressed as

$$\mathbf{G}_{kU}[n] = \mathbf{H}_{kU}[n]\mathbf{F}_{kU}^{1/2}[n], \quad (11)$$

where $\mathbf{H}_{kU}[n] = [h_{pp',mm'}[n]]_{L_U \times L_k}$ is a $L_U \times L_k$ matrix and $\mathbf{F}_{kU}[n] = [\beta_{kU}[n]]_{L_k \times L_k}$ is a $L_k \times L_k$ diagonal matrix. The channel matrices for the other links can be similarly defined by using (11) and properly replacing the indices. However, in the U2R and R2U cases, the channel coefficients are only affected by the movement of ARSU, since GRSU is static.

The achievable rate of the V2U massive MIMO channel can be expressed as

$$r_{kU}[n] = B \sum_{l=1}^{\min(L_k, L_U)} \log_2\left(1 + \frac{p_k^{\text{off}}[n]\lambda_{kU,l}^2[n]}{BN_0 L_k}\right), \quad (12)$$

where $B$ is the allocated bandwidth, $p_k^{\text{off}}[n]$ is the transmit power of the $k$-th vehicle, $N_0$[1] is the variance of the additive white Gaussian noise (AWGN) at ARSU, and $\{\lambda_{kU,l}[n]\}_{l=1}^{\min(L_k,L_U)}$ are the singular values of $\mathbf{G}_{kU}[n]$. Using (12) and appropriately replacing the indices, the achievable rates $r_{UR}$, $r_{RU}$, and $r_{Uk}$ of the U2R, R2U, and U2V massive MIMO channels, respectively, can be obtained.

## III. COMPUTATION OFFLOADING AND DOWNLOADING MODEL

We define $l_k = \{c_k, b_k, \xi_k\}$ the computation task of the $k$-th vehicle, where $c_k > 0$ is the number of required central processing unit (CPU) cycles per bit, $b_k$ is the task-input data size (in bits), $c_k b_k$ is the total required number of CPU cycles, and $\xi_k$ is the proportionality ratio between offloaded data and computed results. The maximum CPU frequency at the $k$-th vehicle and the ARSU is denoted as $f_{k,\max}$ and $f_{U,\max}$, respectively, with $f_{k,\max} < f_{U,\max}$, whereas $c_U > 0$ denotes the number of required CPU cycles per bit at the ARSU. Since the computational resources of the $k$-th vehicle are limited, more computing power is required to accomplish its task within the maximum allowable latency (task deadline) $\eta_k \leq T_U$. In this regard, partial task offloading is exploited. Hence, the $k$-th vehicle offloads to ARSU and GRSU (via relaying) part of its task in each timeslot. The computation task at the $k$-vehicle in a given timeslot is partitioned as

$$b_k[n] = b_{k,l}[n] + b_{k,U}[n] + b_{k,R}[n] \geq b_{k,\min}[n], \quad (13)$$

where $b_{k,l}[n]$, $b_{k,U}[n]$, and $b_{k,R}[n]$ are the computation bits allocated for local computing, offloading to ARSU for computing, and offloading to GRSU for computing via ARSU, respectively. Besides, $b_{k,\min}$ are the minimum task bits that should be periodically completed in each timeslot. Note that the case that $\eta_k = T_U$ is only considered in this paper $\forall k$.

### A. Transmission Delay and Computation Delay

In order to implement the data offloading and downloading processes, while avoiding interference among the vehicles, each timeslot is fairly divided into $K$ equal durations $\{\tau_k[n]\}_{k=1}^{K}$ with $\sum_{k=1}^{K} \tau_k[n] = \tau$. Contrary to previous work on conventional UAV-based MEC networks with single antennas (e.g., [15], [30]), the use of massive MIMO can meaningfully shorten the transmission time thus rendering the duration of data offloading, computing, and downloading comparable. Thus, a protocol for data transmission and data computation with five distinct operation phases is proposed. As shown in Fig. 2, $\tau_k^{\text{off}}[n] = b_{k,UR}[n]/R_{kU}$ is the transmission time for offloading the bits $b_{k,UR}[n] = b_{k,U}[n] + b_{k,R}[n]$ from the $k$-th vehicle to the ARSU (Phase 1); $\tau_{k,U}^{\text{off}}[n] = b_{k,R}[n]/r_{UR}$ is the transmission time for offloading $b_{k,R}[n]$ from ARSU to GRSU via relaying (Phase 2); $\tau_{k,cU}[n] = c_U b_{k,U}[n]/f_{U,\max}$ is the computation delay at the ARSU (Phase 3); $\tau_{k,U}^{\text{dow}}[n] = \tilde{b}_{k,U}[n]/R_{Uk}$ is the transmission time for downloading $\tilde{b}_{k,U}[n] = \xi_k b_{k,U}[n]$ from the ARSU to the $k$-th vehicle (Phase 4); and $\tau_{k,R}^{\text{dow}}[n] = \tilde{b}_{k,R}[n]/R_{Uk}$ is the transmission time for downloading $\tilde{b}_{k,R}[n] = \xi_k b_{k,R}[n]$ from the ARSU to the $k$-th vehicle (Phase 5). Note that the downloading transmission time for the R2U channel is omitted, since there are no transmit power constraints at the GRSU side. Also, the computing time at the GRSU is negligible, whereas the decision time for task partitioning is very short compared to the entire latency and can be neglected as well. It is assumed that the vehicle can simultaneously carry out local computing and bits offloading, whereas the local computation delay $\tau_{k,cl}[n] = c_k b_{k,l}[n]/f_{k,\max}$ can span a timeslot $\tau$. Overall, the following time allocation constraints should be satisfied:

$$0 \leq \left\{\tau_k^{\text{off}}[n], \tau_{k,U}^{\text{off}}[n], \tau_{k,cU}[n], \tau_{k,U}^{\text{dow}}[n], \tau_{k,R}^{\text{dow}}[n], \frac{\tau_{k,cl}[n]}{K}\right\} \leq \frac{\tau}{K}, \quad (14)$$

$$\tau_k^{\text{off}}[n] + \tau_{k,U}^{\text{off}}[n] + \tau_{k,cU}[n] + \tau_{k,U}^{\text{dow}}[n] + \tau_{k,R}^{\text{dow}}[n] \leq \frac{\tau}{K}. \quad (15)$$

As the bits offloading (or bits downloading) cannot outreach the rate capabilities of the massive MIMO channels, we obtain the following constraints:

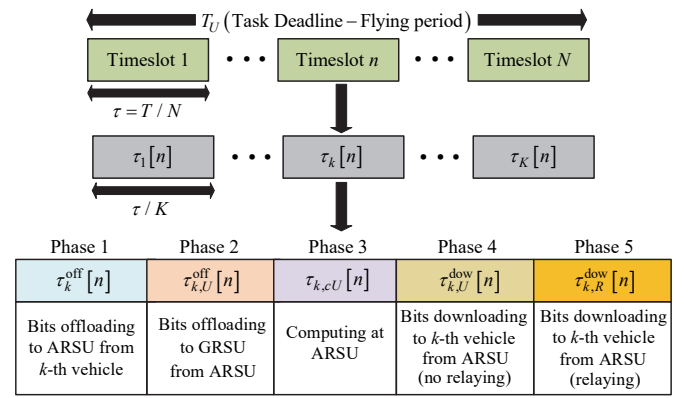

Fig. 2. The data transmission and data computation protocol.

---
[1] Without loss of generality, we assume that $N_0$ is the variance of the AWGN at any network node.

$$b_{k,UR}[n] \leq \tau_k^{\text{off}}[n] r_{kU}[n], \quad (16)$$

$$b_{k,R}[n] \leq \tau_{k,U}^{\text{off}}[n] r_{UR}[n], \quad (17)$$

$$\tilde{b}_{k,U}[n] \leq \tau_{k,U}^{\text{dow}}[n] r_{Uk}[n], \quad (18)$$

$$\tilde{b}_{k,R}[n] \leq \tau_{k,R}^{\text{dow}}[n] r_{Uk}[n]. \quad (19)$$

*B. Energy Consumption*

The energy consumed for data offloading (or computed data downloading) at Phases 1, 2, 4, and 5, respectively, can be expressed as

$$E_k^{\text{off}}[n] = p_k^{\text{off}}[n] \tau_k^{\text{off}}[n] \leq P_{k,\max}^{\text{off}} \tau_k^{\text{off}}[n], \quad (20)$$

$$E_{k,U}^{\text{off}}[n] = p_{k,U}^{\text{off}}[n] \tau_{k,U}^{\text{off}}[n] \leq P_{k,A,\max}^{\text{off}} \tau_{k,U}^{\text{off}}[n], \quad (21)$$

$$E_{k,U}^{\text{dow}}[n] = p_{k,U}^{\text{dow}}[n] \tau_{k,U}^{\text{dow}}[n] \leq P_{k,U,\max}^{\text{dow}} \tau_{k,U}^{\text{dow}}[n], \quad (22)$$

$$E_{k,R}^{\text{dow}}[n] = p_{k,R}^{\text{dow}}[n] \tau_{k,R}^{\text{dow}}[n] \leq P_{k,R,\max}^{\text{dow}} \tau_{k,R}^{\text{dow}}[n], \quad (23)$$

where $p_{k,U}^{\text{off}}[n]$, $p_{k,U}^{\text{dow}}[n]$, and $p_{k,R}^{\text{dow}}[n]$ are the transmit powers in Phases 2, 4, and 5, respectively, and $P_{k,\max}^{\text{off}}$, $P_{k,U,\max}^{\text{off}}$, $P_{k,U,\max}^{\text{dow}}$, and $P_{k,R,\max}^{\text{dow}}$ are the maximum transmit powers at Phases 1, 2, 4, and, 5, respectively. In each timeslot, the energy consumption for local intra-vehicle and ARSU computing can be, respectively, expressed as [50]

$$E_{k,cl}[n] = P_{k,cl} \tau_{k,cl}[n] \equiv \kappa_k c_k^3 (b_{k,l}[n])^3 \tau^{-2}, \quad (24)$$

$$E_{k,cU}[n] = P_{k,cU} \tau_{k,cU}[n] \equiv \kappa_U c_U^3 K^2 (b_{k,U}[n])^3 \tau^{-2}, \quad (25)$$

where $P_{k,cl} = \kappa_k f_{k,\max}^3$ and $P_{k,cU} = \kappa_U f_{U,\max}^3$ is the power consumption of the CPU at the *k*-th vehicle and at the ARSU, respectively [50], and $\kappa_k > 0$ and $\kappa_U > 0$ are the effective capacitance coefficient that rely on the chip architecture at the *k*-th vehicle and at the ARSU, respectively.

## IV. PROBLEM FORMULATION AND OPTIMIZATION

Considering a dual-MEC UAV-aided massive MIMO vehicular network, a novel multi-variable optimization problem is formulated to minimize the WTEC. This problem is explicitly subjected to physical layer parameters, such as the transmit power allocation, as well as the massive MIMO uplink and downlink data rates. Also, this problem accounts for timeslot scheduling and task allocation. Towards this end, the optimization problem is formulated as

$$(P1): \min_{\mathbf{B},\mathbf{P},\boldsymbol{\tau}} \sum_{n=1}^{N} \left( \left( \sum_{k=1}^{K} w_k E_k[n] \right) + w_U E_U[n] \right) \quad (26a)$$

$$\text{s.t. } b_{k,l}[n] + b_{k,U}[n] + b_{k,R}[n] \geq b_{k,\min}[n] \quad (26b)$$

$$b_{k,l}[n] \geq 0, \ b_{k,U}[n] \geq 0, \ b_{k,R}[n] \geq 0 \quad (26c)$$

$$0 \leq \tau_k^{\text{off}}[n] \leq \frac{\tau}{K}, \ 0 \leq \tau_{k,U}^{\text{off}}[n] \leq \frac{\tau}{K}, \ 0 \leq \frac{c_U b_{k,U}[n]}{f_{U,\max}} \leq \frac{\tau}{K},$$

$$0 \leq \tau_{k,U}^{\text{dow}}[n] \leq \frac{\tau}{K}, \ 0 \leq \tau_{k,R}^{\text{dow}}[n] \leq \frac{\tau}{K}, \ 0 \leq \frac{c_k b_{k,l}[n]}{f_{k,\max}} \leq \tau \quad (26d)$$

$$\tau_k^{\text{off}}[n] + \tau_{k,U}^{\text{off}}[n] + \frac{c_U b_{k,U}[n]}{f_{U,\max}} + \tau_{k,U}^{\text{dow}}[n] + \tau_{k,R}^{\text{dow}}[n] \leq \frac{\tau}{K} \quad (26e)$$

$$b_{k,U}[n] + b_{k,R}[n] \leq \tau_k^{\text{off}}[n] r_{kU}\left(\frac{E_k^{\text{off}}[n]}{\tau_k^{\text{off}}[n]}\right) \quad (26f)$$

$$b_{k,R}[n] \leq \tau_{k,U}^{\text{off}}[n] r_{UR}\left(\frac{E_{k,U}^{\text{off}}[n]}{\tau_{k,U}^{\text{off}}[n]}\right) \quad (26g)$$

$$\xi_k b_{k,U}[n] \leq \tau_{k,U}^{\text{dow}}[n] r_{Uk}\left(\frac{E_{k,U}^{\text{dow}}[n]}{\tau_{k,U}^{\text{dow}}[n]}\right) \quad (26h)$$

$$\xi_k b_{k,R}[n] \leq \tau_{k,R}^{\text{dow}}[n] r_{Uk}\left(\frac{E_{k,R}^{\text{dow}}[n]}{\tau_{k,R}^{\text{dow}}[n]}\right) \quad (26i)$$

$$0 \leq E_k^{\text{off}}[n] = p_k^{\text{off}}[n] \tau_k^{\text{off}}[n] \leq P_{k,\max}^{\text{off}} \tau_k^{\text{off}}[n] \quad (26j)$$

$$0 \leq E_{k,U}^{\text{off}}[n] = p_{k,U}^{\text{off}}[n] \tau_{k,U}^{\text{off}}[n] \leq P_{k,A,\max}^{\text{off}} \tau_{k,U}^{\text{off}}[n] \quad (26k)$$

$$0 \leq E_{k,U}^{\text{dow}}[n] = p_{k,U}^{\text{dow}}[n] \tau_{k,U}^{\text{dow}}[n] \leq P_{k,U,\max}^{\text{dow}} \tau_{k,U}^{\text{dow}}[n] \quad (26l)$$

$$0 \leq E_{k,R}^{\text{dow}}[n] = p_{k,R}^{\text{dow}}[n] \tau_{k,R}^{\text{dow}}[n] \leq P_{k,R,\max}^{\text{dow}} \tau_{k,R}^{\text{dow}}[n] \quad (26m)$$

where

$$E_k[n] = E_{k,cl}[n] + E_k^{\text{off}}[n]$$
$$= \kappa_k c_k^3 (b_{k,l}[n])^3 \tau^{-2} + p_k^{\text{off}}[n] \tau_k^{\text{off}}[n], \quad (27)$$

$$E_U[n] = \sum_{k=1}^{K} \left( E_{k,U}^{\text{off}}[n] + E_{k,cU}[n] + E_{k,U}^{\text{dow}}[n] + E_{k,R}^{\text{dow}}[n] \right)$$
$$= \sum_{k=1}^{K} \left( p_{k,U}^{\text{off}}[n] \tau_{k,U}^{\text{off}}[n] + \kappa_U c_U^3 K^2 (b_{k,U}[n])^3 \tau^{-2} \right.$$
$$\left. + p_{k,U}^{\text{dow}}[n] \tau_{k,U}^{\text{dow}}[n] + p_{k,R}^{\text{dow}}[n] \tau_{k,R}^{\text{dow}}[n] \right) \quad (28)$$

are the energy consumption of the *k*-vehicle and ARSU, respectively, in each timeslot, including the energy consumed for bits offloading/downloading and bits computation, $\mathbf{B} \triangleq \{b_{k,l}[n], b_{k,U}[n], b_{k,R}[n]\}$, $\mathbf{P} \triangleq \{p_k^{\text{off}}[n], p_{k,U}^{\text{off}}[n], p_{k,U}^{\text{dow}}[n], p_{k,R}^{\text{dow}}[n]\}$, and $\boldsymbol{\tau} \triangleq \{\tau_k^{\text{off}}[n], \tau_{k,U}^{\text{off}}[n], \tau_{k,U}^{\text{dow}}[n], \tau_{k,R}^{\text{dow}}[n]\}$ are the optimizing variables, and $w_k \geq 0$ and $w_U \geq 0$ are the weight factors of energy consumption of *k*-th vehicle and ARSU, respectively. Depending on the requirements of an envisioned application, the weight factors can be easily modified to satisfy the energy demands and trade-offs. Also, these weight factors provide priority and fairness among the vehicles. As $w_k (w_U)$ increases, the *k*-th vehicle (ARSU) has higher priority and thus the minimization of the *k*-th vehicle's (ARSU's) energy consumption becomes more important. One observes that $E_{total}$ in problem (P1) is an increasing function of the offloaded data. Hence, offloading should be realized, only if local computation violates latency constraints.

*Lemma 1:* Problem (P1) is a convex problem.

*Proof:* From (26a) (27), and (28), we can conclude that the objective function of problem (P1) is convex with respect to (w.r.t.) **P**, $b_{k,l}[n]$, and $b_{k,U}[n]$, since its Hessian matrix is positive semidefinite. Also, the expressions in constraints (26b)-(26e) and (26j)-(26m) are linear. Moreover, the right-hand-side of (26f)-(26i) are concave, since $f(x,t) = t \log(1 + x/t)$ with $t > 0$ is concave [51]. Therefore, Problem (P1) is a convex problem.

To solve Problem (P1), the Lagrangian dual method is leveraged. The optimal solutions of problem (P1) are given in the following proposition.

*Proposition 1:* The optimal computation bits allocated for local computing, offloading to ARSU for computing, and offloading to GRSU for computing via ARSU can be obtained, respectively, as

$$b_{k,l}^*[n] = \tau \left[ \sqrt{\frac{\chi_{1,k,n}}{3 w_k \kappa_k c_k^3}} \right]_0^{\frac{f_{k,\max}}{c_k}}, \quad (29)$$

$$b_{k,U}^*[n] = \begin{cases} \frac{\tau}{K} \left[ \sqrt{\frac{\zeta_{k,n}}{3 w_U \kappa_U c_U^3 f_{U,\max}}} \right]_0^{\frac{f_{U,\max}}{c_U}}, & \zeta_{k,n} \geq 0 \\ 0, & \zeta_{k,n} < 0 \end{cases} \quad (30)$$

$$b_{k,R}^*[n] = \begin{cases} 0, & \chi_{3,k,n} + \chi_{4,k,n} + \chi_{6,k,n}\xi_k - \chi_{1,k,n} > 0 \\ \omega, & \chi_{3,k,n} + \chi_{4,k,n} + \chi_{6,k,n}\xi_k - \chi_{1,k,n} = 0 \end{cases}, \quad (31)$$

where $\zeta_{k,n} = f_{U,\max}(\chi_{1,k,n} - \chi_{3,k,n} - \chi_{5,k,n}\xi_k) - \chi_{2,k,n} c_U$, $\omega$ is an arbitrary non-negative constant, and $\chi_{1,k,n}$, $\chi_{2,k,n}$, $\chi_{3,k,n}$, $\chi_{4,k,n}$, $\chi_{5,k,n}$, and $\chi_{6,k,n}$ are the non-negative Lagrange multipliers (dual variables) related with the constraints in (26b), (26e), (26f), (26g), (26h), and (26i), respectively. Also, the optimal transmit power $p_k^{\text{off}*}[n]$ at Phase 1 can be obtained by solving the following equation using numerical solving techniques:

$$w_k - \chi_{3,k,n} B \sum_{l=1}^{\min(L_k, L_U)} \frac{\lambda_{kU,l}^2[n]}{BN_0 L_k \log_2\left(1 + \frac{p_k^{\text{off}}[n] \lambda_{kU,l}^2[n]}{BN_0 L_k}\right)} = 0. \quad (32)$$

Moreover, the optimal transmission delay $\tau_k^{\text{off}*}[n]$ can be expressed as in (33), shown at the top of the next page page.

*Proof:* See Appendix A.

Using (29) and (30), the optimal computation delays $\tau_{k,cl}^*[n] = c_k b_{k,l}^*[n] / f_{k,\max}$ and $\tau_{k,cU}^*[n] = c_U b_{k,U}^*[n] / f_{U,\max}$ can be obtained. Also, the optimal energy consumption of the k-th vehicle for data offloading at Phase 1 can be expressed as

$$E_k^{\text{off}*}[n] = p_k^{\text{off}*}[n] \tau_k^{\text{off}*}[n]. \quad (34)$$

In addition, using (32)-(34) and properly replacing the indices, the optimal transmit powers, transmission delays, and energy consumption at Phases 2, 4, and 5 can be obtained.

*Remark 1:* The expressions in (29) and (30) indicate that the weight factor directly controls the division of the task-input bits and the corresponding computation delay. As $w_k$ and $w_U$ increase, less task-input data is processed locally and at ARSU, respectively. Also, $b_{k,U}^*[n]$ decreases, as $\xi_k$ increases. Moreover, the k-th vehicle (ARSU) would choose to offload data to ARSU (GRSU) for computing, as far as $b_k[n] > b_{k,l}^*[n]$ ($b_{k,UR}[n] > b_{k,U}^*[n]$).

To derive closed-form solutions for $p_k^{\text{off}*}[n]$ and provide network design recommendations, the special, but common, case of highly correlated rank-1 LoS massive MIMO channels is initially studied. This case leads to the lower bound of the achievable rate [41], [45]. Then, $\mathbf{H}_{kU}[n]$ has one non-zero singular value. Next, the upper bound of the achievable rate is investigated, where this matrix is full-rank and all of its singular values are nonzero and equal. Unlike in Rayleigh channels, orthogonality among the spatially multiplexed signals can be attained in this special case, under strict geometrical constraints and specific orientation of the arrays [43]. Although it is infeasible to adjust the placement of URPAs in highly mobile vehicular scenarios, we investigate the special case of full-rank channels, since it corresponds to the theoretical upper bound of the achievable rate.

*Proposition 2:* The optimal transmit power and offloading time at Phase 1 for the lower bound of the achievable rate can be, respectively, obtained as

$$p_{k,lb}^{\text{off}*}[n] = \left[ B\left( \frac{\chi_{3,k,n}}{w_k \ln(2)} - \frac{N_0 L_k}{\Phi_{kU}[n]} \right) \right]_0^{p_{k,\max}^{\text{off}}} \quad (35a)$$

and as in (35b), shown at the top of the next page, where $\Phi_{kU}[n] = tr\left( \mathbf{G}_{kU}[n] \mathbf{G}_{kU}^H[n] \right) = \sum_{l=1}^{\min(L_k, L_U)} \lambda_{kU,l}^2[n]$ and $(\cdot)^H$ and $tr(\cdot)$ represent the conjugate transpose and trace of matrix, respectively. Moreover, the optimal transmit power and offloading time at Phase 1 for the upper bound of the achievable rate can be, respectively, obtained as

$$p_{k,ub}^{\text{off}*}[n] = \left[ \min(L_k, L_U) p_{k,lb}^{\text{off}*}[n] \right]_0^{p_{k,\max}^{\text{off}}} \quad (36a)$$

and as in (36b), shown at the top of the next page.

*Proof:* See Appendix B.

$$t_k^{\text{off}*}[n] = \begin{cases} = \dfrac{\tau}{K}, & w_k p_k^{\text{off}*}[n] + \chi_{2,k,n} - \chi_{3,k,n} B \sum_{l=1}^{\min(L_k, L_U)} \log_2\left(1 + \dfrac{p_k^{\text{off}*}[n] \lambda_{kU,l}^2[n]}{BN_0 L_k}\right) < 0 \\ \in \left[0, \dfrac{\tau}{K}\right], & w_k p_k^{\text{off}*}[n] + \chi_{2,k,n} - \chi_{3,k,n} B \sum_{l=1}^{\min(L_k, L_U)} \log_2\left(1 + \dfrac{p_k^{\text{off}*}[n] \lambda_{kU,l}^2[n]}{BN_0 L_k}\right) = 0 \\ = 0, & w_k p_k^{\text{off}*}[n] + \chi_{2,k,n} - \chi_{3,k,n} B \sum_{l=1}^{\min(L_k, L_U)} \log_2\left(1 + \dfrac{p_k^{\text{off}*}[n] \lambda_{kU,l}^2[n]}{BN_0 L_k}\right) > 0 \end{cases} \quad (33)$$

$$\tau_{k,lb}^{\text{off}*}[n] = \begin{cases} = \dfrac{\tau}{K}, & w_k p_{k,lb}^{\text{off}*}[n] + \chi_{2,k,n} - \chi_{3,k,n} B \log_2\left(1 + \dfrac{p_{k,lb}^{\text{off}*}[n] \Phi_{kU}[n]}{BN_0 L_k}\right) < 0 \\ \in \left[0, \dfrac{\tau}{K}\right], & w_k p_{k,lb}^{\text{off}*}[n] + \chi_{2,k,n} - \chi_{3,k,n} B \log_2\left(1 + \dfrac{p_{k,lb}^{\text{off}*}[n] \Phi_{kU}[n]}{BN_0 L_k}\right) = 0 \\ = 0, & w_k p_{k,lb}^{\text{off}*}[n] + \chi_{2,k,n} - \chi_{3,k,n} B \log_2\left(1 + \dfrac{p_{k,lb}^{\text{off}*}[n] \Phi_{kU}[n]}{BN_0 L_k}\right) > 0 \end{cases} \quad (35b)$$

$$\tau_{k,ub}^{\text{off}*}[n] = \begin{cases} = \dfrac{\tau}{K}, & w_k p_{k,ub}^{\text{off}*}[n] + \chi_{2,k,n} - \chi_{3,k,n} B \min(L_k, L_U) \log_2\left(1 + \dfrac{p_{k,ub}^{\text{off}*}[n] \Phi_{kU}[n]}{BN_0 L_k \min(L_k, L_U)}\right) < 0 \\ \in \left[0, \dfrac{\tau}{K}\right], & w_k p_{k,ub}^{\text{off}*}[n] + \chi_{2,k,n} - \chi_{3,k,n} B \min(L_k, L_U) \log_2\left(1 + \dfrac{p_{k,ub}^{\text{off}*}[n] \Phi_{kU}[n]}{BN_0 L_k \min(L_k, L_U)}\right) = 0 \\ = 0, & w_k p_{k,ub}^{\text{off}*}[n] + \chi_{2,k,n} - \chi_{3,k,n} B \min(L_k, L_U) \log_2\left(1 + \dfrac{p_{k,ub}^{\text{off}*}[n] \Phi_{kU}[n]}{BN_0 L_k \min(L_k, L_U)}\right) > 0 \end{cases} \quad (36b)$$

---

*Remark 2:* From (35a) and (36a), one concludes that there exists a linear relationship between the optimized transmitted power of a rank-1 and that of a full-rank channel, whereas $p_{k,lb}^{\text{off}*}[n] \leq p_{k,ub}^{\text{off}*}[n]$. As $\min(L_k, L_U)$ increases, $p_{k,ub}^{\text{off}*}[n]$ linearly increases and leads to increased WTEC. Clearly, the massive MIMO channels and system geometry directly affect the energy optimization. In addition, (35a) and (36a) indicate that larger values of $w_k$ correspond to decreased optimized transmit power and thus decreased $E_k^{\text{off}*}[n]$.

Henceforth, predetermined dual variables are considered. The next step of the optimization procedure is to obtain the optimal dual variables by following the procedure described in Appendix C. Since $\boldsymbol{\tau}^*$ and $b_{k,R}^*[n]$ are not unique, we formulate the following linear programming problem:

$$(P2): \min_{b_{k,R}[n], \boldsymbol{\tau}} \sum_{n=1}^{N} \left[ \left( \sum_{k=1}^{K} w_k E_k^{\text{off}}[n] \right) + w_U E_U'[n] \right] \quad (37a)$$

s.t. (26c), (26d), (26e), (26j)-(26m) $\quad (37b)$

$$b_{k,l}^*[n] + b_{k,U}^*[n] + b_{k,R}[n] \geq b_{k,\min}[n] \quad (37c)$$

$$b_{k,U}^*[n] + b_{k,R}[n] \leq \tau_k^{\text{off}}[n] r_{kU}\left(p_k^{\text{off}*}[n]\right) \quad (37d)$$

$$b_{k,R}[n] \leq \tau_{k,U}^{\text{off}}[n] r_{UR}\left(p_{k,U}^{\text{off}*}[n]\right) \quad (37e)$$

$$b_{k,U}^*[n] \xi_k \leq \tau_{k,U}^{\text{dow}}[n] r_{Uk}\left(p_{k,U}^{\text{dow}*}[n]\right) \quad (37f)$$

$$b_{k,R}[n] \xi_k \leq \tau_{k,R}^{\text{dow}}[n] r_{Uk}\left(p_{k,R}^{\text{dow}*}[n]\right) \quad (37g)$$

where $E_U'[n] = \sum_{k=1}^{K}\left(E_{k,U}^{\text{off}}[n] + E_{k,U}^{\text{dow}}[n] + E_{k,R}^{\text{dow}}[n]\right)$. Thus, problem (P2) should be solved, in order to obtain the optimal solution to primal problem (P1). Based on the previous results and observations, the subgradient-based Algorithm 1 is proposed to optimally solve this problem. The complexity and running time of Algorithm 1 depends on the number of time slots and the number of vehicles. More importantly, the main complexity of Algorithm 1 lies in steps 4, 5, and 6, where the complexity is $O(KN)$, $O(KN)$, and $O(K^2 N^2)$, respectively [51]. Hence, Algorithm 1 has a total complexity of $O(K^4 N^4)$. Finally, in Step 9, the complexity mainly depends on solving problem (P2) by CVX [52].

## V. NUMERICAL RESULTS AND DISCUSSION

In this section, numerical results are presented to illustrate the TCCD $\tau_{TCCD} = \sum_{n=1}^{N} \sum_{k=1}^{K} \tau_k[n]$ and the WTEC for different values of the key system parameters and under latency constraints. The effectiveness of the massive MIMO and the optimization method is also studied, whereas the convergence performance of the proposed algorithm is also evaluated. The results take into account the number of antennas, the number of vehicles, the computation task size, the relative location of the ARSU w.r.t. the vehicles (GRSU), the time horizon $T_U$, the velocity and weight factor of energy consumption of ARSU, and the proportionality ratio between offloaded data and computed results. The vector $\mathbf{B}_V \in \mathbb{R}^{1 \times k}$ is used to represent the set of required computation data (in Mbits), in which the $k$-th entry stands for the required computation task for the $k$-th vehicle per timeslot. Without loss of generality, it is assumed that $L_k = L_U = L_R$, whereas the vehicles have identical task requirement. Unless otherwise stated, the values of key parameters are listed in Table I.

Fig. 3 shows the non-optimized and optimized TCCD of Phases 1-5 as a function of the number of vehicles for a rotary-wing ARSU, URPAs with different number of antenna elements, and task requirement $b_k[n] = 0.5$ Mbit per time slot. Clearly, the delay substantially decreases with the number of antennas and grows with the number of vehicles. Besides, the number of supported vehicles changes with the number of antenna elements. To provide computing services to three vehicles, while satisfying the stringent latency constraints, URPAs with at least 16-elements are required. Meanwhile, the optimized scheme supports a larger number of vehicles, when compared with the non-optimized one, thus revealing the effectiveness of our optimization method.

**Algorithm 1** Optimal Solution to Problem (P1)

1: **Set** $K$, $\{w_k\}$, $w_U$, $T_U$, $\tau$, $f_{k,\max}$, $f_{U,\max}$, $\{c_k\}$, $c_U$, $\{\kappa_k\}$, $\kappa_U$, $\{\xi_k\}$, $h_U$, $\{\theta_k\}$, $\theta_R$, $\{v_k\}$, $\{\gamma_k\}$, $v_U$, $\gamma_{U,xy}$, $\gamma_{U,z}$, $c_1$, $c_2$, $c_3$, $v_{tip}$, $v_0$, $d_r$, $s$, $\rho$, $G$, $P_0$, $P_1$, $P_2$, $\lambda$, $\alpha$, $B$, $\beta_0$, $N_0$, $\{P_{k,\max}^{\text{off}}\}$, $\{P_{k,U,\max}^{\text{off}}\}$, $\{P_{k,U,\max}^{\text{dow}}\}$, $\{P_{k,R,\max}^{\text{dow}}\}$, $\{L_k\}$, $L_U$, $L_R$, $\{\delta_k\}$, $\delta_U$, $\delta_R$, $\beta_x$, $\beta_y$, $\beta_z$, and the tolerant threshold $\varepsilon$.

2: **Initialize** the iteration index, the non-optimized dual variables $\{\mathbf{x}_\delta\}_{\delta=1}^6$ and the ellipsoid (as described in Appendix C). Then, obtain the channel matrices using (11) and decompose these matrices via singular value decomposition (SVD) to obtain the singular values.

3: **Repeat**

4: Use (29) and (30) and obtain $b_{k,l}^*[n]$ and $b_{k,U}^*[n]$, respectively. Then, numerically solve the equation in (32) and obtain $\mathbf{P}^*$. Also, use (33) and obtain $\boldsymbol{\tau}^*$. Calculate the WTEC.

5: Solve problem P1-dual defined in (A.3a)-(A.3c) in Appendix A by calculating the subgradients defined in (C.1a)-(C1.f) in Appendix C.

6: Update $\{\mathbf{x}_\delta\}_{\delta=1}^6$ according to the ellipsoid method.

7: **End Repeat** until convergence.

8: Let $\{\mathbf{x}_\delta^*\}_{\delta=1}^6 \leftarrow \{\mathbf{x}_\delta\}_{\delta=1}^6$

9: Use (29) and (30) and obtain $b_{k,l}^*[n]$ and $b_{k,U}^*[n]$, respectively. Update $\mathbf{P}^*$ by re-solving the equation in (32). Then, obtain $b_{k,R}^*[n]$ and $\boldsymbol{\tau}^*$ by solving problem (P2) by CVX. Finally, obtain the minimum WTEC.

TABLE I
DEFINITION, NOTATION, AND VALUES OF KEY PARAMETERS

| System Parameters | Value |
|---|---|
| Number of vehicles: $K$ | 3 |
| Weight factor for energy consumption for $k$-th vehicle (ARSU): $w_k$ ($w_U$) | 1 (0.1) |
| Parameters of fixed-wing ARSU: $c_1$, $c_2$, $c_3$ | $9.26 \cdot 10^{-4}$, 2250, 3.33 [26] |
| Parameters of rotary-wing ARSU: $v_{tip}$, $v_0$, $d_r$, $s$, $\rho$, $G$, $P_0$, $P_1$, $P_2$ | 120, 4.3, 0.6, 0.05, 1.225, 0.503, $12 \cdot 30^3 \cdot 0.4^3 \rho s G / 8$, $1.1 \cdot 20^{3/2} / \sqrt{2\rho G}$, 11.46 [26] |
| **Computation Parameters** | **Value** |
| Task deadline (flight duration of ARSU): $T_U$ | 8 s |
| Timeslot length: $\tau$ | 0.2 s [30] |
| Maximum CPU frequency at $k$-th vehicle (ARSU): $f_{k,\max}$ ($f_{U,\max}$) | 1 GHz (3 GHz) |
| Required CPU cycles per bit at $k$-th vehicle (ARSU): $c_k$ ($c_U$) | $10^3$ ($10^3$) cycles/bit [30] |
| CPU capacitance coefficient at $k$-th vehicle (ARSU): $\kappa_k$ ($\kappa_U$) | $10^{-27}$ ($10^{-27}$) [30] |
| Task size ratio of output data to input data: $\xi_k$ | 0.8 |
| **Geometrical and Mobility Parameters** | **Value** |
| Initial height of ARSU: $h_U$ | 10 m [15] |
| Initial elevation angle of ARSU relative to $O_1$, $O_2$, $O_3$, $O_R$, respectively: $\theta_1$, $\theta_2$, $\theta_3$, $\theta_R$ | $\pi/3$, $\pi/4$, $\pi/6$, $\pi/3$ |
| Slant, downtilt, and bearing angle, respectively: $\beta_x$, $\beta_y$, $\beta_z$ | $\pi/3$, $\pi/4$, $\pi/3$ |
| Velocity and moving direction of $k$-th vehicle in the azimuth domain, respectively: $v_k$, $\gamma_k$ | 60 km/h, $\pi/3$ |
| Velocity and moving direction of ARSU in the azimuth (elevation) domain, respectively: $v_U$, $\gamma_{U,xy}$ ($\gamma_{U,z}$) | 10 m/s, $\pi/3$ ($\pi/9$) |
| **Wireless Transmission Parameters** | **Value** |
| Carrier wavelength: $\lambda$ | 0.15 m |
| Path-loss exponent: $\alpha$ | 2 [49] |
| Bandwidth for uplink (or downlink): $B$ | 5 MHz |
| Channel gain at reference distance $d_0 = 1$ m: $\beta_0$ | -50 dB [30] |
| Variance of AWGN at $k$-th vehicle, ARSU, and GRSU: $N_0$ | -130 dBm/Hz [30] |
| Max. transmit power in Phase 1, 2, 4, and, 5, respectively: $P_{k,\max}^{\text{off}}$, $P_{k,U,\max}^{\text{off}}$, $P_{k,U,\max}^{\text{dow}}$, $P_{k,R,\max}^{\text{dow}}$ | 35 dBm [30] |
| Number of antennas at $k$-th vehicle, ARSU, and GRSU array, respectively: $L_k$, $L_U$, $L_R$ | 36, 36, 36 |
| Inter-element spacing at $k$-th vehicle, ARSU, and GRSU antenna array, respectively: $\delta_k$, $\delta_U$, $\delta_R$ | $\lambda/2$ |

Fig. 4 shows the non-optimized and optimized TCCD as a function of the number of antenna elements for a rotary-wing ARSU and varying task requirement $\mathbf{B}_V$. One observes that the TCCD significantly decreases, as the number of antennas increases, owing to the higher data rates and the lower transmission delay. As the number of antennas increases from 16 to 64, up to 1 Mbits and 1.55 Mbits can be supported for the non-optimized and optimized scheme, respectively. Also, using URPAS with relatively small dimensions, e.g., 36-

element URPAs of 0.45 m x 0.45 m size for the commonly used 2 GHz carrier frequency, up to 0.82 Mbits can be computed per time slot. Besides, using a similar setup and single-antennas less than 0.2 Mbits can be timely executed. Thus, the benefits and feasibility from integrating large-scale antennas on size-constrained conventional vehicles and currently available commercial off-the-shelf UAVs is affirmed. Overall, a reasonable number of antennas should be employed, according to the amount of offloaded data, in order to attain acceptable TCCD, while satisfying practical antenna size constraints. By using mmWave frequency bands, which are potentially available for air-to-ground-communications [53], the antenna arrays can be even more compact and more demanding tasks can be handled.

Fig. 5 investigates the impact of the altitude of a hovering rotary-wing ARSU on the non-optimized WTEC for $K=1$, $b_1[n]=0.6$ Mbits, and varying initial elevation angle of the ARSU w.r.t. the vehicle, i.e., $\theta_1$, and w.r.t. the GRSU, i.e., $\theta_R$. As $h_U$ increases, the ARSU draws away from both the vehicle and GRSU and more energy is consumed. It is also evident that the WTEC fairly increases as $\theta_1$ decreases, since the quality of the V2U and U2V channels in terms of the path-loss and correlation is somehow degraded. However, changing $\theta_R$ is even less influential and negligibly affects the WTEC. Previous results on single-antenna configurations [15], stated that the UAV should be closed to ground nodes to ensure low offloading/downloading energy consumption and support large task sizes. Nevertheless, in this paper, the use of massive MIMO promises meaningfully lower transmission delays and enhanced rates that obliterate such indications and compensate the increased path-loss observed in larger distances. Therefore, this paper suggests that the ARSU should not necessarily approach the moving vehicles and/or GRSU to attain satisfactory WTEC and/or meliorate possible side effects of unstable $h_U$ due to obstacles and wind/pressure variableness. By avoiding aimless movements, a significant amount of propulsion energy can be saved thus extending the endurance of the ARSU.

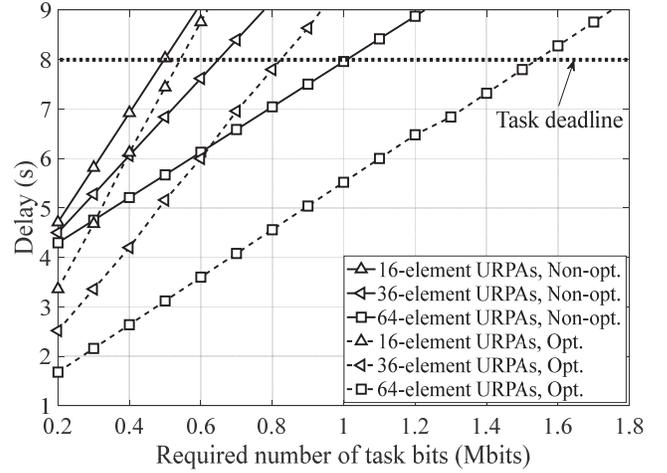

Fig. 4. The non-optimized and optimized TCCD as a function of the task requirement per time slot for varying number of antennas.

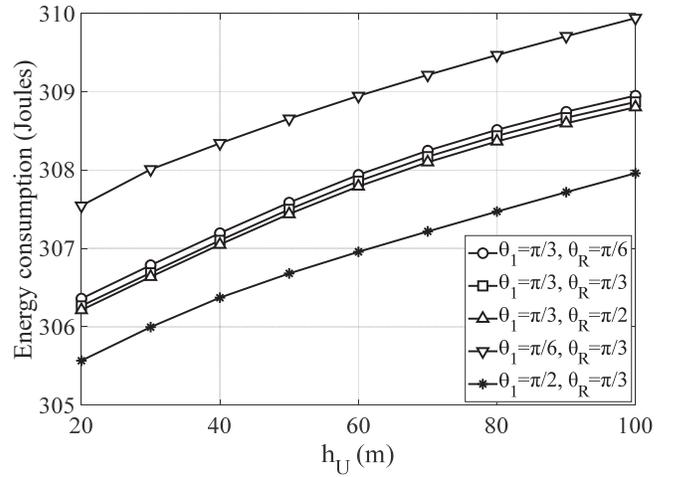

Fig. 5. The non-optimized WTEC as a function of the altitude of the ARSU for varying elevation angle of the ARSU.

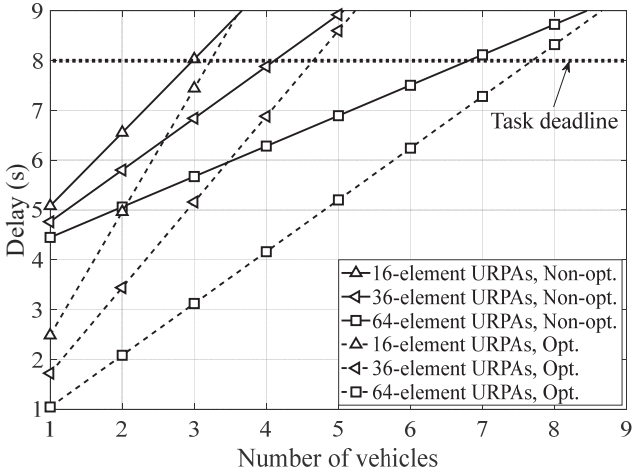

Fig. 3. The non-optimized and optimized TCCD as a function of the number of vehicles for varying number of antennas.

Figs. 6 illustrates the non-optimized and optimized WTEC as a function of the task requirement per time slot for a fixed-wing ARSU and six computing scenarios, including local computing, full offloading and partial offloading. Apparently, the GRSU is not necessary to assist on computation for small values of task bits, e.g., 0.1 Mbits per time slot. However, local computing is subject to a maximum computing capability $\tau f_{k,\max} / c_k$, according to (14). Exploiting only the ARSU for computing leads to even greater WTEC, since wireless transmission consumes additional energy. Besides, as indicated in (25), the energy consumption for computing at ARSU exponentially increases with $K$. In order to extend the supported number of task bits, the vehicles and the ARSU may cooperatively handle the computation process with an acceptable growth of the WTEC. On the other hand, demanding computation tasks presuppose the participation of the GRSU for efficient edge computing and slight energy cost. As the task bits increase the vehicles should tend to transmit ideally the entire amount of task bits to the GRSU via the

ARSU. Clearly, the curves of the optimized schemes outperform the non-optimized ones and depict the advantages of partial offloading by capitalizing on the local computation resources, as well as the MEC resources at ARSU and GRSU. Also, the difference between the non-optimized and optimized schemes enlarges with the task requirement, while the common scenario of rank-1 channels constitutes the most energy-efficient solution.

Fig. 7 shows the curves of the optimized WSEC as a function of the task completion time (flying period) for a fixed-wing ARSU for varying $v_U$, $w_U$, and $\xi_U$, when the task requirement is $\mathbf{B}_V = [b_1, b_2, b_3] = [0.6, 0.6, 0.6]$ Mbits and $\xi_1 = \xi_2 = \xi_3$. One observes that the consumed energy drastically and linearly increases, as the stringent deadline increases. It is also obvious that WTEC increases as $v_U$ and $w_U$ step up, since the propulsion energy contributes more to the WTEC. Highly-intensive computation tasks, e.g., video-rendering applications and delivery of 360º videos, may lead to $\xi_k \gg 1$ [54]. However, owing to the enhanced spectral efficiency offered by the massive MIMO channels, changing $\xi_k$ negligibly affects the WTEC. This is not the case for conventional single-antenna scenarios [15].

Finally, Fig. 8 investigates the convergence efficiency of the proposed Algorithm 1 and demonstrates the optimized WTEC for a fixed-wing ARSU and tolerant threshold $\varepsilon = 10^{-4}$ as a function of the iteration index. It can be seen that the proposed optimized WTEC scheme nearly converges after about 7 iterations, regardless of the task sizes and number of antennas, thus achieving computational effectiveness.

## VI. CONCLUSIONS AND FUTURE DIRECTIONS

In this paper, a novel WTEC optimization problem for a delay-constrained massive MIMO UAV-aided MEC-enabled vehicular network has been formulated. This problem can be decomposed into multiple convex subproblems that can be solved by the Lagrangian dual method along with an efficient subgradient-based algorithm. Capitalizing on the convenient form of the closed-form solutions, numerical calculations have been carried out to illustrate the mathematical derivations. We showed that the number of antennas determines the number of supported vehicles and the size of offloaded data, under latency constraints. It has been also demonstrated that the vehicles may perform local computation for low task requirements. As task bits increase, partial task offloading is necessary. Since the velocity and weight factor of ARSU control the propulsion energy, the proposed approach has underlined that massive MIMO can counterbalance the distance-dependent path-loss and reduce purposeless mobility of ARSU. This work can be expanded into various fertile research areas. As a pre-determined ARSU's trajectory is considered, the 3-D trajectory optimization is envisioned as a future work. Multiple ARSUs and GRSUs can be utilized along with learning-based methods for intelligent control, in order to extend the network range. Also, massive connectivity can be ensured by adopting NOMA, while mm-wave frequencies can further increase the array gain.

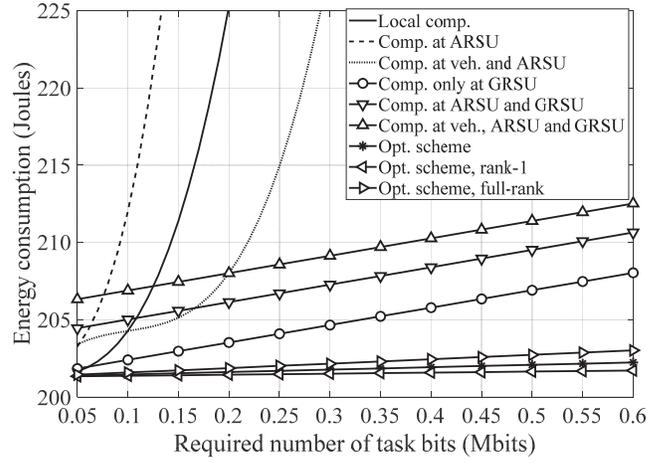

Fig. 6. The non-optimized and optimized WTEC as a function of the task requirement per time slot for different computing scenarios.

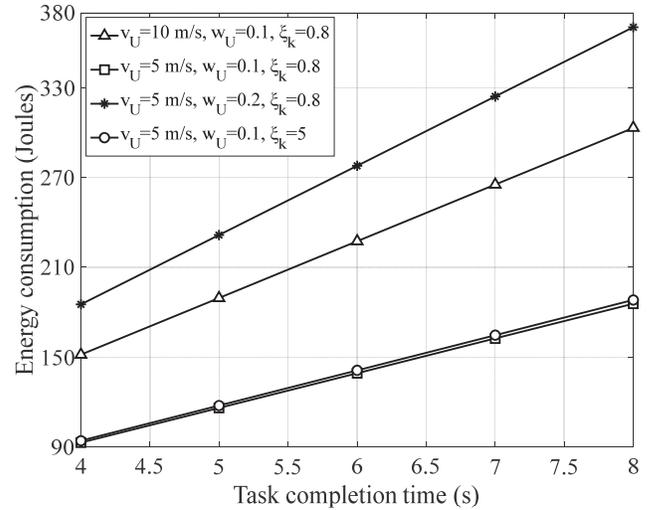

Fig. 7. The optimized WTEC as a function of the task completion time for varying velocity of the ARSU, weight factor of the ARSU, and task size ratio of output data to input data.

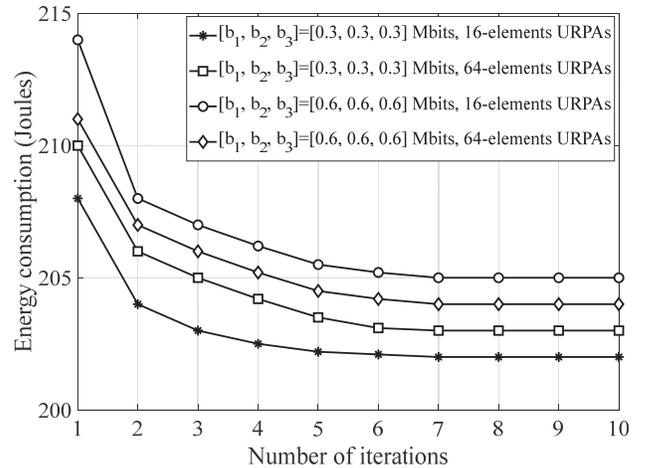

Fig. 8. The optimized WTEC as a function of the number of iterations for varying task requirement and number of antenna elements.

# APPENDIX A
## PROOF OF PROPOSITION 1

The Lagrange function of problem (P1) can be expressed as in (A.1), shown at the top of the next page, where $\mathbf{x}_1$, $\mathbf{x}_2$, $\mathbf{x}_3$, $\mathbf{x}_4$, $\mathbf{x}_5$, and $\mathbf{x}_6$ denote the sets of the dual variables $\chi_{1,k,n}$, $\chi_{2,k,n}$, $\chi_{3,k,n}$, $\chi_{4,k,n}$, $\chi_{5,k,n}$, and $\chi_{6,k,n}$, respectively. For arbitrary dual variables, the dual function of problem (P1) can be written as

$$\xi(\mathbf{x}_1,\mathbf{x}_2,\mathbf{x}_3,\mathbf{x}_4,\mathbf{x}_5,\mathbf{x}_6) = \min_{\mathbf{B},\mathbf{P},\tau} \mathcal{L}(\mathbf{B},\mathbf{P},\tau,\mathbf{x}_1,\mathbf{x}_2,\mathbf{x}_3,\mathbf{x}_4,\mathbf{x}_5,\mathbf{x}_6)$$
(A.2a)

$$\text{s.t. (26c), (26d), (26j)-(26m)} \quad \text{(A.2b)}$$

Based on the results in [30], $\xi(\mathbf{x}_1,\mathbf{x}_2,\mathbf{x}_3,\mathbf{x}_4,\mathbf{x}_5,\mathbf{x}_6)$ is bounded, provided that the inequality $\chi_{3,k,n} + \chi_{4,k,n} + \chi_{6,k,n}\xi_k - \chi_{1,k,n} \geq 0$ holds. Thus, the dual problem of problem (P1) can be expressed as

P1-dual: $\max_{\mathbf{x}_1,\mathbf{x}_2,\mathbf{x}_3,\mathbf{x}_4,\mathbf{x}_5,\mathbf{x}_6} \xi(\mathbf{x}_1,\mathbf{x}_2,\mathbf{x}_3,\mathbf{x}_4,\mathbf{x}_5,\mathbf{x}_6)$ (A.3a)

$$\text{s.t. } \{\mathbf{x}_1,\mathbf{x}_2,\mathbf{x}_3,\mathbf{x}_4,\mathbf{x}_5,\mathbf{x}_6\} \succeq 0 \quad \text{(A.3b)}$$

$$\chi_{3,k,n} + \chi_{4,k,n} + \chi_{6,k,n}\xi_k - \chi_{1,k,n} \geq 0 \quad \text{(A.3c)}$$

Since problem (P1) is convex, the Slater's condition is satisfied [51]. Owing to the strong duality between (P1) and (P1-dual), we obtain the optimal solution of problem (P1) by solving its dual problem, i.e., problem (P1-dual). For arbitrary values of $\{\mathbf{x}_\delta\}_{\delta=1}^6$, we obtain the dual function by solving the problem defined in (A.3). This problem can be rewritten into a set of $KN$ independent subproblems. Thus, we can further decompose these subproblems into several subproblems w.r.t. each vehicle. These subproblems are convex and their solutions satisfy the Karush–Kuhn–Tucker (KKT) conditions [51].

We define the subproblems related with $b_{k,l}[n]$, $b_{k,U}[n]$, and $b_{k,R}[n]$, respectively as

(L1): $\min_{b_{k,l}[n]} w_k \kappa_k c_k^3 (b_{k,l}[n])^3 \tau^{-2} - \chi_{1,k,n} b_{k,l}[n]$

s.t. (26c), (26d)

(L2): $\min_{b_{k,U}[n]} w_U \kappa_U c_U^3 K^2 (b_{k,U}[n])^3 \tau^{-2}$

$+ \left( \frac{\chi_{2,k,n} c_U}{f_{U,\max}} + \chi_{3,k,n} + \chi_{5,k,n}\xi_k - \chi_{1,k,n} \right) b_{k,U}[n]$

s.t. (26c), (26d)

(L3): $\min_{b_{k,R}[n]} (\chi_{3,k,n} + \chi_{4,k,n} + \chi_{6,k,n}\xi_k - \chi_{1,k,n}) b_{k,R}[n]$

s.t. (26c)

By solving subproblems (L1), (L2), and (L3) with the aid of KKT conditions, we obtain the optimal solutions in (29), (30), and (31), respectively. Also, we define the subproblem, which is related with $p_k^{\text{off}}[n]$ and $\tau_k^{\text{off}}[n]$, as

(L4): $\min_{\tau_k^{\text{off}}[n], p_k^{\text{off}}[n]} \left( w_k p_k^{\text{off}}[n] + \chi_{2,k,n} \right) \tau_k^{\text{off}}[n]$

$- \chi_{3,k,n} \tau_k^{\text{off}}[n] B_0 \sum_{l=1}^{\min(L_k,L_U)} \log_2 \left( 1 + \frac{p_k^{\text{off}}[n] \lambda_{kU,l}^2[n]}{B_0 N_0 L_k} \right)$

s.t. (26d), (26j)

The Lagrangian of subproblem (L4) can be written as

$\mathcal{L}_4(\psi_{1,k,n},\psi_{2,k,n},\psi_{3,k,n},\psi_{4,k,n}) = \left( w_k p_k^{\text{off}}[n] + \chi_{2,k,n} \right) \tau_k^{\text{off}}[n]$

$- \chi_{3,k,n} \tau_k^{\text{off}}[n] B_0 \sum_{l=1}^{\min(L_k,L_U)} \log_2 \left( 1 + \frac{p_k^{\text{off}}[n] \lambda_{kU,l}^2[n]}{B_0 N_0 L_k} \right)$

$- \psi_{1,k,n} \tau_k^{\text{off}}[n] - \psi_{2,k,n}\left( \tau - \tau_k^{\text{off}}[n] \right) - \psi_{3,k,n} p_k^{\text{off}}[n] \tau_k^{\text{off}}[n]$

$- \psi_{4,k,n}\left( P_{k,\max}^{\text{off}} \tau_k^{\text{off}}[n] - p_k^{\text{off}}[n] \tau_k^{\text{off}}[n] \right),$ (A.4)

where $\psi_{1,k,n}$, $\psi_{2,k,n}$, $\psi_{3,k,n}$, and $\psi_{4,k,n}$ are non-negative Lagrange multipliers associated with the constraints $\tau_k^{\text{off}}[n] \geq 0$, $\tau_k^{\text{off}}[n] \leq \tau/K$, $p_k^{\text{off}}[n]\tau_k^{\text{off}}[n] \geq 0$, and $p_k^{\text{off}}[n]\tau_k^{\text{off}}[n] \leq P_{k,\max}^{\text{off}}\tau_k^{\text{off}}[n]$, respectively, specified in (26d) and (26j). Based on KKT, the optimal transmit power at Phase 1 can be obtained by solving the equation $\partial \mathcal{L}_4(\psi_{1,k,n},\psi_{2,k,n},\psi_{3,k,n},\psi_{4,k,n}) / \partial p_k^{\text{off}}[n] = 0$, which is defined in (32), using numerical solving techniques. Then, $\tau_k^{\text{off}*}[n]$ can be obtained by substituting $p_k^{\text{off}*}[n]$ into subproblem (L4) and is expressed as in (33).

# APPENDIX B
## PROOF OF PROPOSITION 2

The lower bound of the achievable rate can be expressed as [41], [45]

$$r_{kU,lb}[n] = B \log_2 \left( 1 + \frac{p_k^{\text{off}}[n] \Phi_{kU}[n]}{B N_0 L_k} \right). \quad \text{(B.1)}$$

Using (B.1) instead of (12), and solving the equation $\partial \mathcal{L}_4(\psi_{1,k,n},\psi_{2,k,n},\psi_{3,k,n},\psi_{4,k,n}) / \partial p_k^{\text{off}}[n] = 0$, we obtain the optimal solution in (35a). Then, the optimal solution in (35b) can be obtained by substituting $p_{k,lb}^{\text{off}*}[n]$ into subproblem (L4). Also, the upper bound of the achievable rate is expressed as [41], [45]

$$r_{kU,ub}[n] = B \min(L_k,L_U) \log_2 \left( 1 + \frac{p_k^{\text{off}}[n] \Phi_{kU}[n]}{B N_0 L_k \min(L_k,L_U)} \right). \quad \text{(B.2)}$$

Similarly, using (B.2), one can obtain (36a) and then (36b).

$$\mathcal{L}(\mathbf{B},\mathbf{P},\boldsymbol{\tau},\mathbf{x}_1,\mathbf{x}_2,\mathbf{x}_3,\mathbf{x}_4,\mathbf{x}_5,\mathbf{x}_6) = \sum_{n=1}^{N}\sum_{k=1}^{K}\left[w_k\kappa_k c_k^3\left(b_{k,l}[n]\right)^3\tau^{-2} + w_k p_k^{\text{off}}[n]\tau_k^{\text{off}}[n]\right]$$

$$+w_U\sum_{n=1}^{N}\sum_{k=1}^{K}\left[p_{k,U}^{\text{off}}[n]\tau_{k,U}^{\text{off}}[n] + \kappa_U c_U^3 K^2\left(b_{k,U}[n]\right)^3\tau^{-2} + p_{k,U}^{\text{dow}}[n]\tau_{k,U}^{\text{dow}}[n] + p_{k,R}^{\text{dow}}[n]\tau_{k,R}^{\text{dow}}[n]\right)\right]$$

$$+\sum_{n=1}^{N}\sum_{k=1}^{K}\chi_{1,k,n}b_{k,\min}[n] - \sum_{n=1}^{N}\sum_{k=1}^{K}\chi_{1,k,n}b_{k,l}[n] + \sum_{n=1}^{N}\sum_{k=1}^{K}\left(\frac{\chi_{2,k,n}c_U}{f_{U,\max}} + \chi_{3,k,n} + \chi_{5,k,n}\xi_k - \chi_{1,k,n}\right)b_{k,U}[n] - \sum_{n=1}^{N}\sum_{k=1}^{K}\frac{\chi_{2,k,n}\tau}{K}$$

$$+\sum_{n=1}^{N}\sum_{k=1}^{K}\left(\chi_{3,k,n} + \chi_{4,k,n} + \chi_{6,k,n}\xi_k - \chi_{1,k,n}\right)b_{k,R}[n] + \sum_{n=1}^{N}\sum_{k=1}^{K}\chi_{2,k,n}\tau_{k,U}^{\text{dow}}[n] + \sum_{n=1}^{N}\sum_{k=1}^{K}\chi_{2,k,n}\tau_k^{\text{off}}[n] + \sum_{n=1}^{N}\sum_{k=1}^{K}\chi_{2,k,n}\tau_{k,U}^{\text{off}}[n]$$

$$+\sum_{n=1}^{N}\sum_{k=1}^{K}\chi_{2,k,n}\tau_{k,R}^{\text{dow}}[n] - \sum_{n=1}^{N}\sum_{k=1}^{K}\chi_{3,k,n}\tau_k^{\text{off}}[n]B\sum_{l=1}^{\min(L_k,L_U)}\log_2\left(1 + \frac{p_k^{\text{off}}[n]\lambda_{kU,l}^2[n]}{BN_0 L_k}\right) - \sum_{n=1}^{N}\sum_{k=1}^{K}\chi_{4,k,n}\tau_{k,U}^{\text{off}}[n]B\sum_{l=1}^{\min(L_U,L_R)}\log_2\left(1 + \frac{p_{k,U}^{\text{off}}[n]\lambda_{UR,l}^2[n]}{BN_0 L_U}\right)$$

$$-\sum_{n=1}^{N}\sum_{k=1}^{K}\chi_{5,k,n}\tau_{k,U}^{\text{dow}}[n]B\sum_{l=1}^{\min(L_U,L_k)}\log_2\left(1 + \frac{p_{k,U}^{\text{dow}}[n]\lambda_{Uk,l}^2[n]}{BN_0 L_U}\right) - \sum_{n=1}^{N}\sum_{k=1}^{K}\chi_{6,k,n}\tau_{k,R}^{\text{dow}}[n]B\sum_{l=1}^{\min(L_U,L_k)}\log_2\left(1 + \frac{p_{k,R}^{\text{dow}}[n]\lambda_{Uk,l}^2[n]}{BN_0 L_U}\right) \quad \text{(A.1)}$$

## APPENDIX C
## DERIVATION OF OPTIMAL DUAL VARIABLES

In order to obtain the optimal dual variables, the problem (P1-dual), which is defined in (A3), should be solved. This problem is convex but non-differentiable and can be iteratively solved using the subgradient ellipsoid method [51]. The convergence of the ellipsoid method is guaranteed by the convexity of problem (P1-dual) [51]. Let $\Delta\mathbf{x}$ be the subgradient of the objective function in (A.3) w.r.t. $\mathbf{x}$. Then, we obtain

$$\Delta\mathbf{x}_1 = b_{k,U}[n] + b_{k,R}[n] - \tau_k^{\text{off}}[n]r_{kU}\left(\frac{E_k^{\text{off}}[n]}{\tau_k^{\text{off}}[n]}\right), \quad \text{(C.1a)}$$

$$\Delta\mathbf{x}_2 = b_{k,R}[n] - \tau_{k,U}^{\text{off}}[n]r_{UR}\left(\frac{E_{k,U}^{\text{off}}[n]}{\tau_{k,U}^{\text{off}}[n]}\right), \quad \text{(C.1b)}$$

$$\Delta\mathbf{x}_3 = \xi_k b_{k,U}[n] - \tau_{k,U}^{\text{dow}}[n]r_{Uk}\left(\frac{E_{k,U}^{\text{dow}}[n]}{\tau_{k,U}^{\text{dow}}[n]}\right), \quad \text{(C.1c)}$$

$$\Delta\mathbf{x}_4 = \xi_k b_{k,R}[n] - \tau_{k,R}^{\text{dow}}[n]r_{Uk}\left(\frac{E_{k,R}^{\text{dow}}[n]}{\tau_{k,R}^{\text{dow}}[n]}\right), \quad \text{(C.1d)}$$

$$\Delta\mathbf{x}_5 = b_{k,\min}[n] - b_{k,l}[n] - b_{k,U}[n] - b_{k,R}[n], \quad \text{(C.1e)}$$

$$\Delta\mathbf{x}_6 = \tau_k^{\text{off}}[n] + \tau_{k,U}^{\text{off}}[n] + \frac{c_U b_{k,U}[n]}{f_{U,\max}} + \tau_{k,U}^{\text{dow}}[n] + \tau_{k,R}^{\text{dow}}[n] - \tau/K. \quad \text{(C.1f)}$$